# Estimating hourly population distribution change at high spatiotemporal resolution in urban areas using geo-tagged tweets, land use data, and dasymetric maps

Ming-Hsiang Tsou[a]*, Hao Zhang[a], Atsushi Nara[a], Su Yeon Han[a]

[a]*The Center for Human Dynamics in the Mobile Age, San Diego State University, USA*

mtsou@mail.sdsu.edu, zhanghaoshogo@gmail.com, anara@mail.sdsu.edu, suhanmappingideas@gmail.com

Abstract:

This paper introduces a spatiotemporal analysis framework for estimating hourly changing population distribution in urban areas using geo-tagged tweets (the messages containing users' physical locations), land use data, and dasymetric maps. We collected geo-tagged social media (tweets) within the County of San Diego during one year (2015) by using Twitter's Streaming Application Programming Interfaces (APIs). A semi-manual Twitter content verification procedure for data cleaning was applied first to separate tweets created by humans and non-human users (bots). The next step is to calculate the number of unique Twitter users every hour with the two different geographical units: (1) census blocks, and (2) 1km by 1km resolution grids of LandScan. The final step is to estimate actual dynamic population by transforming the numbers of unique Twitter users in each census block or grid into estimated population densities with spatial and temporal variation factors. A temporal factor was based on hourly frequency changes of unique Twitter users within San Diego County, CA. A spatial factor was estimated by using the dasymetric method with land use maps and 2010 census data. Several comparison maps were created to visualize the spatiotemporal pattern changes of dynamic population distribution.

# 1. Introduction

## 1.1 *Human dynamics and population distribution*

The prevailing use of social media and mobile phone data provides a great research opportunity for researchers to map and analyze dynamic human behaviors, communications, and movements (Tsou and Leitner 2013; Deville et al. 2014; Pei et al. 2014; Tsou 2015). People use smartphones, mobile devices, and personal computers to build up their digital life and to leave their digital footprints on the Internet. These human-made digital records provide a foundation for human dynamics research. Human dynamics is a new transdisciplinary research field attracting scientists and researchers from different domains, including complex systems (Barabási 2005), video analysis (Bregler 1997; Wang and Singh 2003), spatial diffusion of events (Issa et al. 2017) and geography (Tsou 2014; Han et al. 2017). One key research question of human dynamics is the dynamic change of population distribution in urban areas. Conventionally, the change of population distribution is estimated from census survey with data sampling and forecasting techniques. Recently, scientists have started to use satellite images (Bhaduri et al. 2007), mobile phone data (Bengtsson et al. 2011; Deville et al. 2014), or vehicle probe data (Hara and Kuwahara 2015) to estimate the dynamic change of population distribution at small area level. One example is to use mobile phone-based call detail records (CDR) to detect spatial and temporal differences in everyday activities among multiple cities (Ahas et al. 2015). Another example is to estimate seasonal, weekends/weekdays, and daily changes in population distribution over multiple timescales with aggregated and anonymized mobile phone data (Deville et al. 2014). In Geographic Information Systems (GIS) and cartographic research, dasymetric mapping methods have been applied to estimate population density using census data and ancillary data sources (Wright 1936; Eicher and Brewer 2001; Holt et al. 2004). In the previous studies, the authors have identified that it is a challenging problem to integrate vector-based census tracks and raster-based land cover data and satellite images for dasymetric mapping. To improve the traditional problems of binary value in categorical data and areal weighting, Mennis and Hultgren (2006) introduced an intelligent dasymetric mapping technique (IDM) with a data-driven methodology to calculate the ratio of class densities. Similar to the IDM method, this study utilizes social media data (geo-tagged data), other GIS data sources (land use and census data), and dasymetric mapping techniques to estimate the hourly change of population distribution. There are several advantages of using social media for population estimation. The real-time updates of social media messages can better reflect dynamic changes of population than remote sensing imageries, which are often more expensive in cost and time to collect and process data (Dong et al. 2010). Alternatively, mobile phone data, such as CDR, are also very expensive and inaccessible. Another drawback of CDR is that it is not possible to identify the content of communications in each phone call. In contrast, social media data are easy-to-collect, free (using public access methods), content-rich, and updated in real-time (Tsou 2015; Issa et al. 2017).

**1.2** *Selecting appropriate spatial and temporal scales for population estimation*

This study utilized public Twitter Application Programming Interfaces (APIs) to collect geo-tagged Twitter messages (tweets) through customized Python programs. The geo-tagged tweets were downloaded via the Twitter Streaming APIs and stored in a NoSQL database (MongoDB). We collected geo-tagged tweets within the bounding box of San Diego County for one year (from 2015/1/1 to 2015/12/31). There are total 7884806 geotagged tweets. Among the collected data, 2,601,560 (33.2%) tweets do not contain the exact coordinates and 2,355,945 (30.1%) were created outside the San Diego County. This study only utilized the remaining 2,927,301 (37.7%) geo-tagged tweets within San Diego County for population estimation. We noticed that the number of monthly geo-tagged tweets in San Diego County in 2015 fluctuated. The months of March and April 2015 have highest numbers of geo-tagged tweets. A similar trend reported by other researchers, such as Business Insider (Edward 2016) suspecting that the causes might be due to Twitter's systematic updates. Figure 1 illustrates the spatial distribution of geotagged tweets from 12am to 1am in downtown, San Diego during weekdays in July 2015 (over one month).

Two types of spatial units (U.S. Census blocks and LandScan grids) were selected to estimate the distribution of the population in order to facilitate evacuation planning and emergency management during disaster events in San Diego County. The main reason for choosing census blocks is to match the need for constructing traffic analysis zones (TAZ), which can be aggregated from census blocks. A TAZ is a special area formalized by local transportation officials for analyzing traffic-related data and evacuation planning. The total population within a TAZ zone should not exceed 3000 people. A census block is the smallest geographic unit defined by the U.S. Census Bureau for demographic analysis. Researchers can utilize TAZ to create disaster evacuation plans and emergency response procedures. In this study, the census blocks and their population estimates in San Diego County were obtained from the 2010 Decennial Census data.

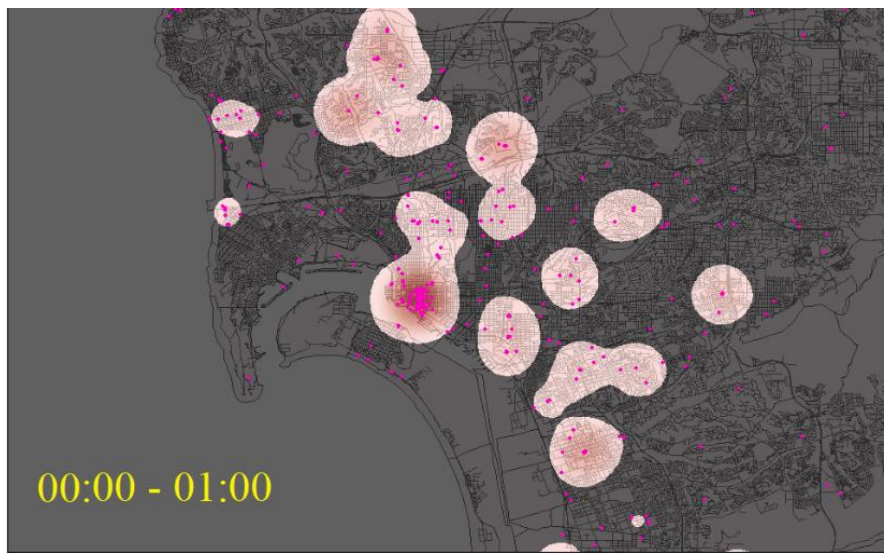

**Figure 1.** The distribution of geo-tagged Twitter messages (tweets as red dots) in San Diego downtown from 12am to 1am during weekdays in July of 2015 (26 days combined).

In addition to census blocks, the geographical unit of LandScan (i.e. the grid having 1km x 1km spatial resolution) was used to compare our result of the estimated population with the LandScan population data for the purpose of model validation. The LandScan grid was developed by the Oak Ridge National Laboratory (Bhaduri et al. 2007).

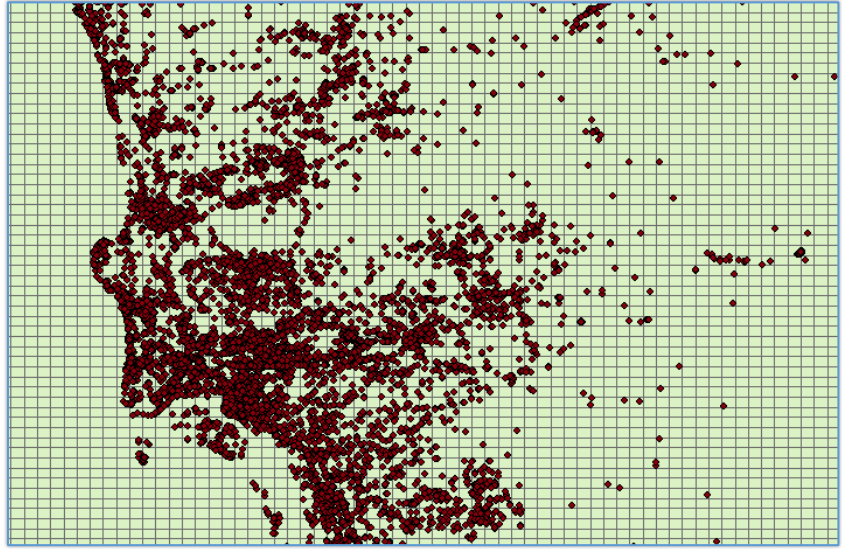

**Figure 2.** The LandScan Grids and the distribution of geo-tagged tweets (red dots) in San Diego County in 2015 (combined 365 days).

We selected one hour as our temporal resolution for estimating population density in San Diego County to meet the need for evacuation planning (Figure 3). During weekdays, the unique Twitter user activities of posting Twitter messages decrease from midnight to 4 am. From 4 am to noon, the user activity starts to climb up. We assume that relatively the large number of Twitter activities around noon is due to tweets related to lunch time activities posted by residents and visitors. The peak of the tweeting activities come at around 8 pm when people are getting dinner or enjoying the leisure time with friends or family members. We also noticed that tweeting activities show different patterns between weekdays (Monday to Friday) and weekends (Saturday and Sunday). In general, the tweeting activities are more active during the weekends comparing to weekdays (Figure 3). Despite the similar pattern found on the weekdays where people tweeted most around 8 pm, the tweeting rate are high at around 2 pm during weekends. Therefore, we distinguish weekdays from weekends for the hourly population density estimation.

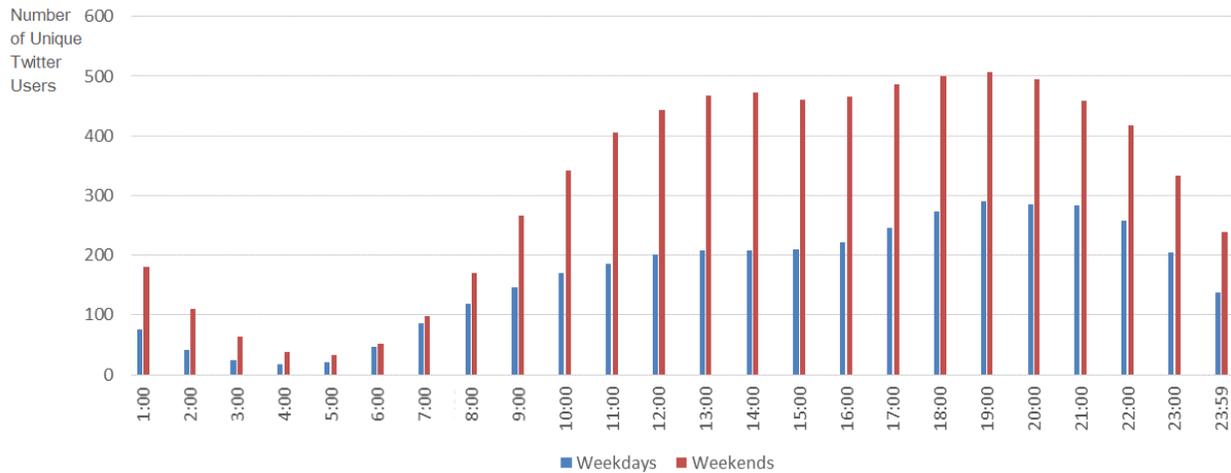

**Figure 3.** Comparison of hourly average numbers of unique Twitter users in San Diego County on weekdays (Monday to Friday) and weekends (Saturday to Sunday) in 2015.

## 2. Data cleaning and pre-processing

### 2.1 *Data cleaning*

Previous research has identified some major types of data noises in Twitter data, including spams, bots, and cyborgs (Yardi et al. 2009; Chu et al. 2012). Spams and bot messages are created for reaching more users and increasing the financial gain for spammers. Since spam and bots messages can not represent the actual locations of human beings, we removed all the identifiable spams and bots based on the source field in Twitter metadata and some general bot detection rules (for example, removing tweets from TweetMyJOBS and others based on a black list of the source field). The major portion of the noise (spams and bots) in San Diego dataset includes job posting (9.07% of the total geo-tagged tweets, such as TweetMyJOBS), advertisements (1.60%, such as dlvr.it), and earthquake (1.06%) in San Diego County. The earthquake event related tweets are geo-tagged in the localities of the earthquakes. In this study, 13.01% of geo-tagged tweets were identified as noises and removed. After removing these spams and bot posts, 2,546,385 tweets were used for calculating the unique Twitter users in each grid or census block within one hour by filtering multiple messages posted by a single user for weekdays and weekends.

### 2.2 *Calculating unique users within census blocks or LandScan grids*

Within each geographical unit of LandScan grids or census blocks, we estimate population during a certain hourly time slot by calculating the frequency of the unique user IDs. Since one Twitter user can post several tweets within an hour from a same region (a census block or a LandScan grid), we counted one unique user ID once within a region during one hour rather than the total number of tweets. Another criterion for population estimation is that each polygon

(census blocks or grids) should have less than 3,000 population in order to match the definition of TAZ. Since the census blocks are created for population analysis, each polygon usually will have less than 3,000 unique users or population. However, some LandScan grid may contain over 3,000 unique users in downtown areas. When one grid contains over 3,000 unique users or population, we need to divide those cell grids into four quadrants. Then, the population distribution can be re-calculated in each sub-cell. When any sub-cell has over 3,000 unique users, we can divide it again. The region quad-tree technique can ensure the decomposition of the densely populated area so that the evacuation model can be applied to both rural (less populated) and urban (more populated) areas without increasing too much on the size of the file on two-dimensional geographic data.

## 3. Dynamic distribution patterns of unique twitter users

### 3.1 *Calculating the hourly unique twitter users in census blocks*

Figure 4 (a) and (b) represent the distribution of unique Twitter users from 6 am – 6:59 am (a) and from 8 pm – 8:59 pm (b) respectively during weekdays in 2015 in San Diego County. The unique Twitter user density was calculated by using the total unique Twitter users within one census block during the specific hour, divided by the area of the census block. Figure 4 (c) displays the 2010 population census data to visually compare its geographical distribution to that of unique Twitter users. In these maps, we selected the quantile classification method at 8pm as the classification framework (applied to other time slots) in order to compare their spatial patterns. Figure 4 (a) and (b) illustrate a significant increase in unique Twitter users from 6 am to 8 pm in Western urbanized areas. The geographical distribution of unique Twitter users from 8 pm – 8:59 pm (Fig. 4 (b)), when has the highest average number of unique Twitter users in 2015 in San Diego County, is similar to that based on the 2010 census data (Fig. 4 (c)).

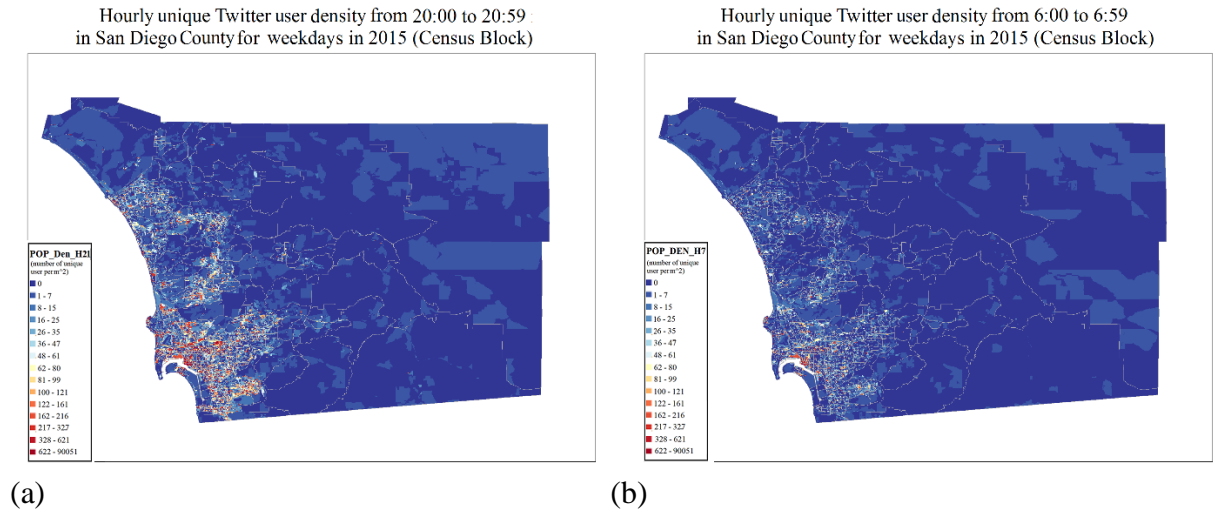

(a)                                                                                            (b)

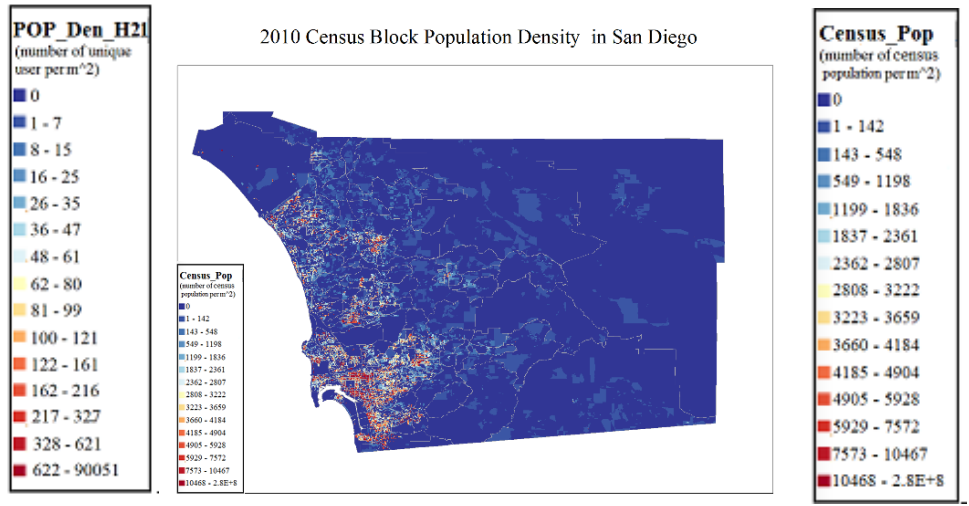

(c)

**Figure 4.** Spatial distribution patterns of unique Twitter users using census blocks in San Diego County from 6am – 6:59am (a) and from 8pm – 8:59pm (b) with 2015 geo-tagged tweets for weekdays. The (c) map displays the population density using 2010 census data.

Maps in Figure5 are enlarged views of Figure 4 exhibiting San Diego City downtown areas. Figure 5 (a) and (b) highlight the increase of the number of unique Twitter users in areas shopping malls in Fashion Valley and Mission Valley, Balboa Park and San Diego Zoo, and the downtown Gaslamp area. The dynamic changes in these areas are reflecting the real world activities in San Diego downtown area. By comparing the 8 pm map (Fig. 5(b)) with the 2010 census block population map (Fig. 5(c)), we found that the large number of unique Twitter users in areas without any population based on the census data. These areas are governmental and commercial lands including Balboa Park, San Diego Zoo, shopping malls, and the San Diego international airport. Since the census population is considered as nighttime population estimated from residential addresses, this example shows the capability of utilizing social media data to estimate population distribution at a finer spatio-temporal scale.

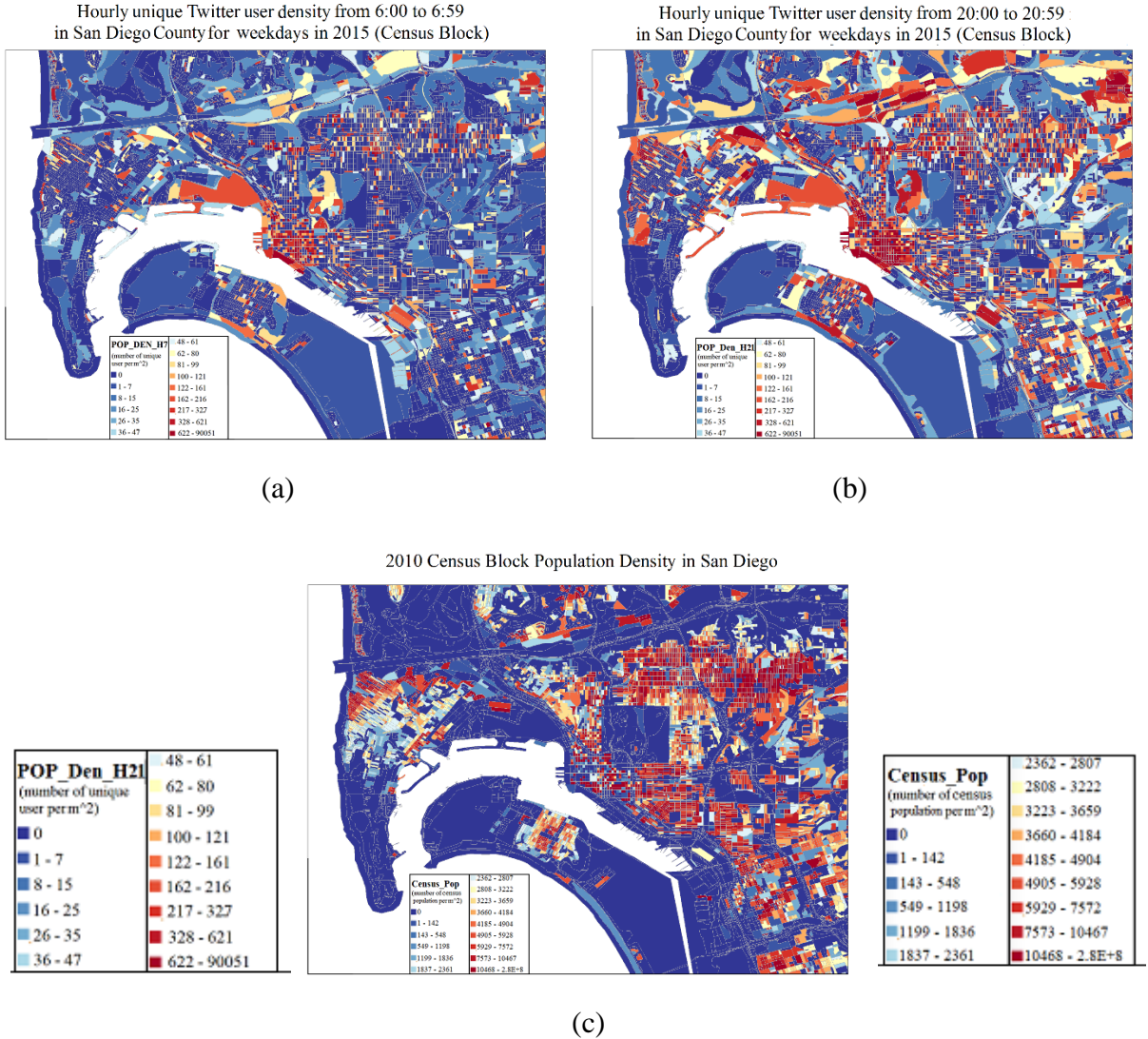

**Figure 5.** The spatial distribution of unique Twitter users in census blocks of San Diego downtown areas from 6 am to 6:59 am (a) and from 8 pm to 8:59 pm (b) with 2015 geo-tagged tweets for weekdays. The (c) map displays the 2010 census data in San Diego downtown areas.

### 3.2 *Calculating the unique twitter users in LandScan grids*

Similar to the comparative study in Section 3.1, we compared the geographical distributions between the unique Twitter users and the LandScan 24-hour population estimate using LandScan grids (Fig. 6 and 7). Similar to the census block maps, the map of 8 pm to 8:59 pm shows a significantly increased the number of unique Twitter users compared to 6 am – 6:59 am in downtown areas and southern part of San Diego City. The 2014 LandScan population pattern (c) seems more enhanced (or exaggerated) comparing to the unique Twitter user population patterns in both 6 am and 8 pm.

We found that one major problem of the LandScan Grids is the low spatial resolution (1km x 1km) comparing to the high spatial resolution of census blocks in urban areas. It is very difficult to manually create quad-tree structure within the selected LandScan Grids (over 3000 people) in order to estimate population in TAZ for evacuation planning. Therefore, our study will only focus on the development of population change models using census blocks in the next sections.

### 3.3 *Comparing the population change patterns of unique twitter users between weekdays and weekends*

With the hourly unique Twitter users density maps being produced (Fig. 4 and Fig. 5) based on weekdays and weekends, some human movement patterns can be detected and further analyzed. One of the advantages of visualizing dynamic Twitter user population patterns is that their dynamic changes can reflect the real world situation with a high spatial resolution (census blocks) and a high temporal resolution (hourly). The following example introduces a case study in the Qualcomm Stadium with the comparison between weekdays and weekends (Fig. 6). The Qualcomm Stadium is a multi-purpose stadium located in San Diego City, CA. The Qualcomm Stadium events data is archived through their official website in the events calendar. During the weekdays, the stadium usually hosts one to three events per day from 15:00 to 20:30. The events held on weekends usually started from 10:30 and ended at 17:30 (Fig. 8). The population density of unique Twitter users in Qualcomm Stadium during the weekdays shows the highest peak of Twitter user activities at 6pm. The high peaks of weekend's activities are from 1pm to 5pm. These patterns match the real world situations since most football game events are happening between 1pm to 5pm on weekends. Figure 7 illustrates the comparison of the unique Twitter user density patterns in the Qualcomm's census blocks between weekday (a) and weekends (b) from 12pm to 12:59pm with its surrounding area. Qualcomm Stadium has a higher density of population at 12pm during weekends (comparing to weekdays).

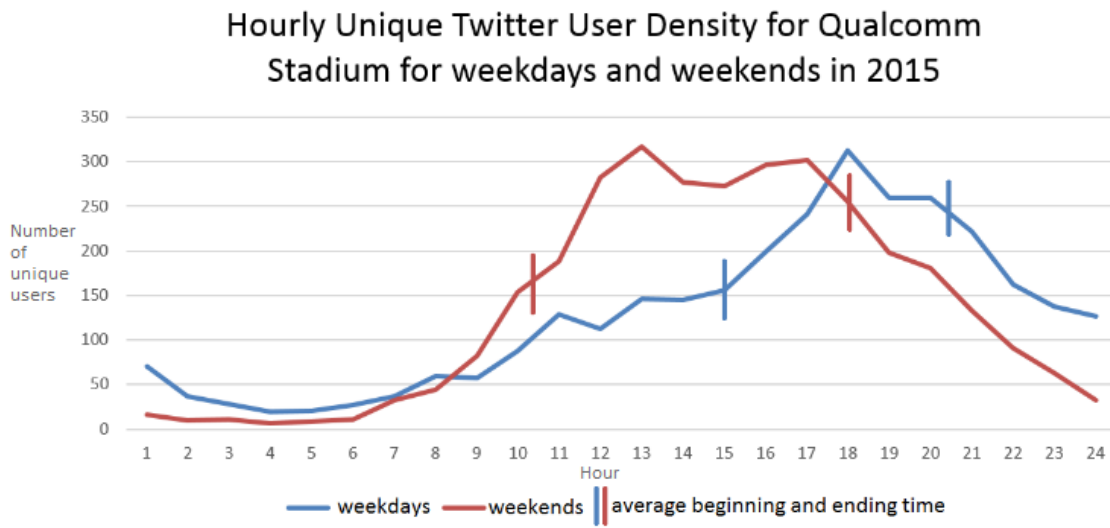

**Figure 6.** Comparing weekdays (blue) and weekends (red) hourly unique Twitter user density in the Qualcomm Stadium census block using 2015 geo-tagged tweets.

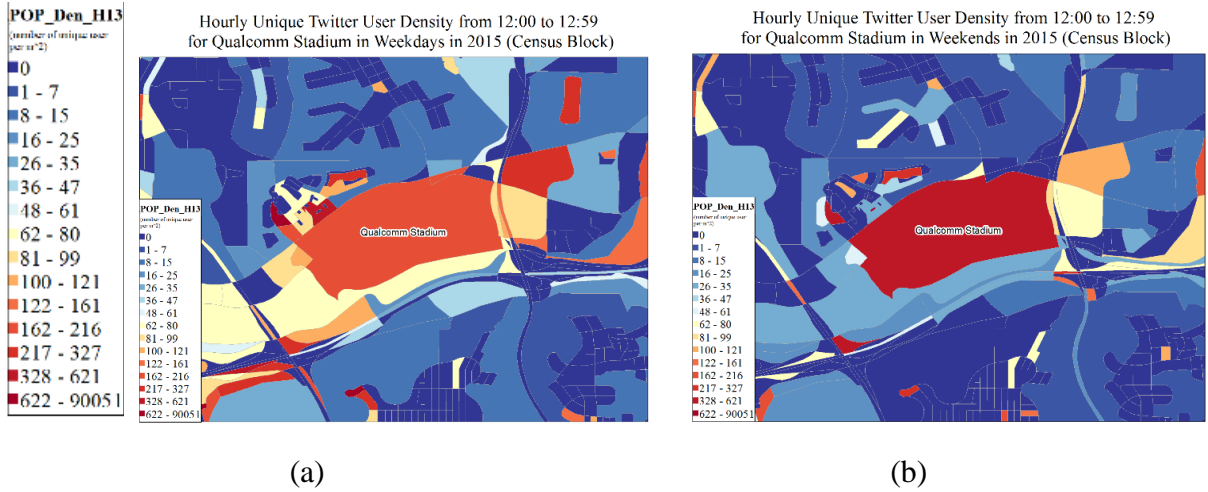

(a) (b)

**Figure 7**. Hourly Unique Twitter User Density from 12pm to 12:59 pm at the Qualcomm Stadium census block for Weekdays (a) and Weekends (b) in 2015.

### 3.4 *Comparing unique twitter population with census and LandScan population*

Comparing the weekdays and weekends unique Twitter user density map in Census Block polygon with census population and LandScan grid with LandScan population can reveal the fact whether Twitter population can be used to represent the human mobility and real human population during different period of time in a day. The Census population represents the population distribution during the nighttime since it collects the number of people living in their household. LandScan population shows an ambient population for the average over 24 hours (Dobson, et al., 2003)

The Table 1 and Figure 8-9 present the $z_{hx \cap pop}$ values in San Diego County area which compares the similarity of census block and LandScan population with unique Twitter user in different time slot from H1 to H24. Each Z value represents for the sum of absolute difference (*SAD* value) of two sets of data within range 0 to 1 based on formula 1.

$$z_{hx \cap pop} = \Sigma \left| \frac{P_{A \cap hx}}{P_{hx_{max}}} - \frac{P_{A \cap pop}}{P_{pop_{max}}} \right| \quad (1)$$

where:

$z_{hx \cap pop}$ = the sum of the absolute difference of number of population between time slot $hx$ and census or LandScan population $pop$;
$P_{A \cap hx}$ = the value of unique Twitter population in time slot $hx$ in Polygon $P_A$;
$P_{hx_{max}}$ = the maximum value of unique Twitter population in time slot $hx$.

Note that *sd* refers to San Diego, *cb* refers to census block polygon, *ls* refers to LandScan grid, *wd* refers weekdays, and *we* refers to weekends. Thus, the intersection between H1 (0:00 to 0:59) and $z_{sd\_cb\_wd}$ stands for the *SAD* Value of comparing the unique Twitter user density map with census block population density in the scale of San Diego County during weekdays. Based on the result show in the table for census block polygon, the H5 (4:00 to 4:49) in weekdays and H6 (5:00 to 5:59) in weekends are the two time slot where the unique Twitter user is the closest to the census block population. The census block population records the number of human population in the residential area in detail. Meanwhile, 4:00 to 5:59 is usually the time when people get up during the morning time. Thus, it is possible to reflect the human residential area by using Twitter data. On the other side, the result of $z_{hx \cap pop}$ value, after comparing the similarity of LandScan population with unique Twitter user show a different result with census data. LandScan data is well known for its effort on providing the ground truth data about human population by combining the census data, light data, slop and topography data into a complex dasymetric mapping model to estimate the population. Thus, by taking the night-time lights into consideration, LandScan data can be considered as one of the dataset which provides the relatively accurate nighttime population distribution data. Based on the result show in the table for LansScan grid, the H24 (23:00 to 23:49) in weekdays and H2 (1:00 to 1:59) in weekends are the two time slot where the unique Twitter user number pattern is the closest to the LandScan population pattern. The result shows the effectiveness of using Twitter data to estimate the night time population based on the LandScan data.

|  |  | San Diego County | | | |
|---|---|---|---|---|---|
|  |  | Census Block Polygons | | LandScan Grids | |
|  |  | Weekdays | Weekends | Weekdays | Weekends |
|  |  | $z_{sd\_cb\_wd}$ | $z_{sd\_cb\_we}$ | $z_{sd\_ls\_wd}$ | $z_{sd\_ls\_we}$ |
| H1 | 0:00 to 0:59 | 402.1 | 412.3 | 37.0 | 18.8 |
| H2 | 1:00 to 1:59 | 399.5 | 403.9 | 35.8 | 18.3 |
| H3 | 2:00 to 2:59 | 430.7 | 408.9 | 38.1 | 25.3 |
| H4 | 3:00 to 3:59 | 377.6 | 402.6 | 51.1 | 32.0 |
| H5 | 4:00 to 4:59 | 366.7 | 367.9 | 33.7 | 31.7 |
| H6 | 5:00 to 5:59 | 367.9 | 367.0 | 42.3 | 30.5 |
| H7 | 6:00 to 6:59 | 381.1 | 377.4 | 73.4 | 43.6 |
| H8 | 7:00 to 7:59 | 387.4 | 386.5 | 56.5 | 48.8 |
| H9 | 8:00 to 8:59 | 391.7 | 381.8 | 50.4 | 43.3 |
| H10 | 9:00 to 9:59 | 390.8 | 388.8 | 51.1 | 44.8 |
| H11 | 10:00 to 10:59 | 391.6 | 388.9 | 45.6 | 42.3 |
| H12 | 11:00 to 11:59 | 391.7 | 397.5 | 41.5 | 44.1 |
| H13 | 12:00 to 12:59 | 393.1 | 396.1 | 41.8 | 45.3 |
| H14 | 13:00 to 13:59 | 394.2 | 399.9 | 41.4 | 46.2 |
| H15 | 14:00 to 14:59 | 392.1 | 398.4 | 42.6 | 45.0 |
| H16 | 15:00 to 15:59 | 392.1 | 396.3 | 44.0 | 42.6 |
| H17 | 16:00 to 16:59 | 398.4 | 398.1 | 42.6 | 43.5 |
| H18 | 17:00 to 17:59 | 411.2 | 392.7 | 42.0 | 44.4 |
| H19 | 18:00 to 18:59 | 387.9 | 396.4 | 42.1 | 38.3 |
| H20 | 19:00 to 19:59 | 390.7 | 392.3 | 41.7 | 37.6 |
| H21 | 20:00 to 20:59 | 405.4 | 394.6 | 42.2 | 34.5 |
| H22 | 21:00 to 21:59 | 428.3 | 397.5 | 39.3 | 32.1 |
| H23 | 22:00 to 22:59 | 441.2 | 392.1 | 36.3 | 28.8 |
| H24 | 23:00 to 23:59 | 428.0 | 409.3 | 31.8 | 25.8 |

*Note: sd = San Diego, cb = Census Block, ls = LandScan, wd = Weedays, we = Weekends

Lowest Value = Most Similar    Highest Value = Most Dissimilar

**Table 1**. The sum of absolute difference between the number of hourly unique twitter data (from 0:00 to 23:59) with census block population and LandScan population during weekdays and weekends in San Diego county.

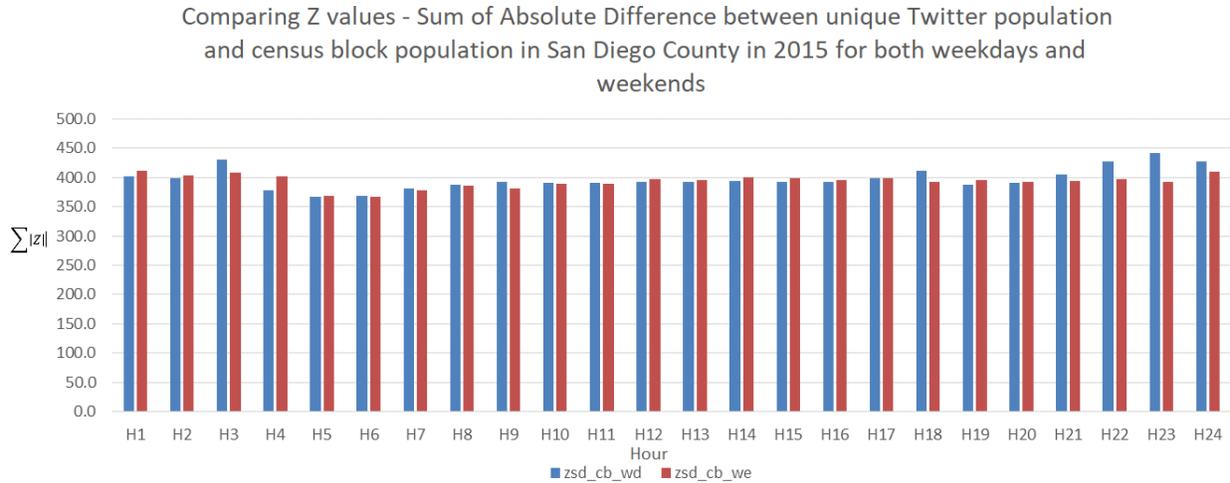

**Figure 8**. Comparing z values – sum of absolute difference between unique twitter population and census block population in San Diego County in 2015 for both weekdays and weekends.

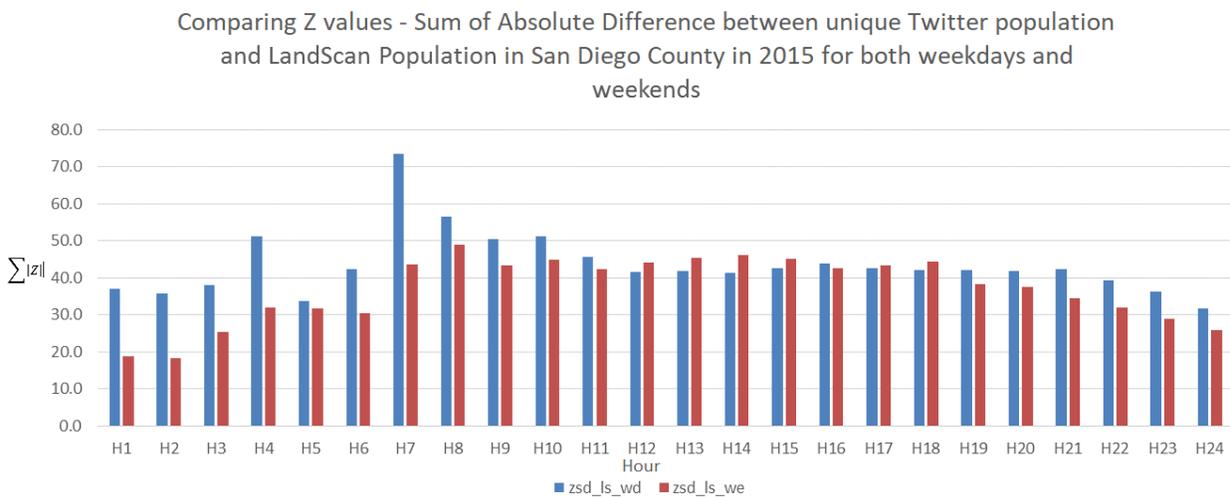

**Figure 9**. Comparing z values – sum of absolute difference between unique twitter population and LandScan population in San Diego county in 2015 for both weekdays and weekends.

Table 2 and Figures 10-11 present the $z_{hx \cap pop}$ values in San Diego County area by comparing the census block population and LandScan population with unique Twitter user in downtown area, San Diego. Note that *dt* refers to downtown area of San Diego, H5 (4:00 to 4:49) for both weekdays and weekends is the time slot where the unique Twitter user is the closest to the census block population. Meanwhile, the result of $z_{hx \cap pop}$ value by comparing the similarity of LandScan population with unique Twitter user show a different result with census data. H24 (23:00 to 23:49) in weekdays and H1 (0:00 to 0:59) in weekends are the two time slot where the unique Twitter user is the closest to the LandScan population. On the other side, from the perspective of dissimilarity, H24 (23:00 to 23:49) and H1 (0:00 to 0:59) have the most dissimilar unique Twitter user distribution comparing to the census block population. In the LandScan, H7 (6:00 to 6:59) and H8 (7:00 to 7:59) are the two time slots which are most dissimilar to the LandScan population density distribution data. It is notable that the time slots that are most similar in census block polygon are close to the most dissimilar in LandScan scale while the most dissimilar in census block polygon are considered as the most similar ones in LandScan scale. The results indicate the difference emphasis for census block dataset and LandScan dataset on how the data was collected and processed.

| | | San Diego County | | | |
| --- | --- | --- | --- | --- | --- |
| | | Census Block Polygons | | LandScan Grids | |
| | | Weekdays | Weekends | Weekdays | Weekends |
| Time Slot | Desciption | $Z_{dt\_cb\_wd}$ | $Z_{dt\_cb\_we}$ | $Z_{dt\_ls\_wd}$ | $Z_{dt\_ls\_we}$ |
| H1 | 0:00 to 0:59 | 131.3 | 120.0 | 11.6 | 5.9 |
| H2 | 1:00 to 1:59 | 126.4 | 116.0 | 11.3 | 6.0 |
| H3 | 2:00 to 2:59 | 121.3 | 116.1 | 12.0 | 8.6 |
| H4 | 3:00 to 3:59 | 109.0 | 113.1 | 16.3 | 10.5 |
| H5 | 4:00 to 4:59 | 97.5 | 98.3 | 11.3 | 11.0 |
| H6 | 5:00 to 5:59 | 97.8 | 98.6 | 14.1 | 11.4 |
| H7 | 6:00 to 6:59 | 101.9 | 102.4 | 23.2 | 15.9 |
| H8 | 7:00 to 7:59 | 104.4 | 106.4 | 18.4 | 17.6 |
| H9 | 8:00 to 8:59 | 106.5 | 105.3 | 17.1 | 16.2 |
| H10 | 9:00 to 9:59 | 106.5 | 108.0 | 17.6 | 16.2 |
| H11 | 10:00 to 10:59 | 107.2 | 108.3 | 15.8 | 15.5 |
| H12 | 11:00 to 11:59 | 107.3 | 111.6 | 14.4 | 16.2 |
| H13 | 12:00 to 12:59 | 108.2 | 111.0 | 14.7 | 16.8 |
| H14 | 13:00 to 13:59 | 108.5 | 112.9 | 14.4 | 16.9 |
| H15 | 14:00 to 14:59 | 108.0 | 112.1 | 14.7 | 16.6 |
| H16 | 15:00 to 15:59 | 108.0 | 111.3 | 15.0 | 15.5 |
| H17 | 16:00 to 16:59 | 110.9 | 112.4 | 14.5 | 15.9 |
| H18 | 17:00 to 17:59 | 116.7 | 110.3 | 14.4 | 16.5 |
| H19 | 18:00 to 18:59 | 107.2 | 111.6 | 14.5 | 13.8 |
| H20 | 19:00 to 19:59 | 108.0 | 110.1 | 14.0 | 13.2 |
| H21 | 20:00 to 20:59 | 114.1 | 110.6 | 13.6 | 11.5 |
| H22 | 21:00 to 21:59 | 123.4 | 111.8 | 12.3 | 10.4 |
| H23 | 22:00 to 22:59 | 129.2 | 110.0 | 11.5 | 9.4 |
| H24 | 23:00 to 23:59 | 132.8 | 118.6 | 10.2 | 8.5 |

*Note: dt = Downtown, cb = Census Block, ls = LandScan, wd = Weedays, we = Weekends
Lowest Value = Most Similar    Highest Value = Most Dissimilar

**Table 2**. Sum of absolute difference between the number of hourly unique twitter data (from 0:00 to 23:59) with census block population and LandScan population during weekdays and weekends in San Diego downtown area.

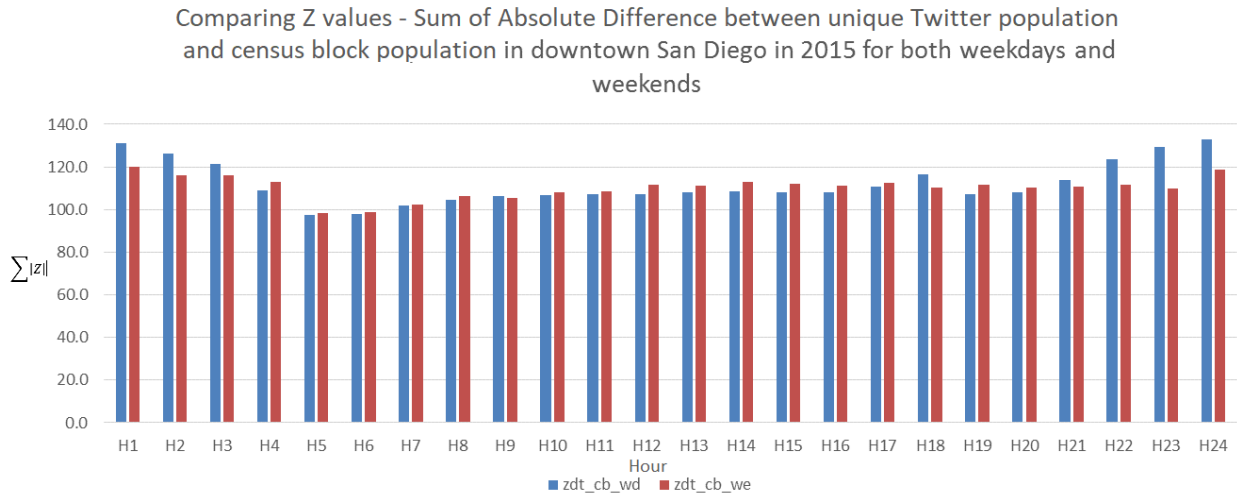

**Figure 10**. Comparing Z values – Sum of Absolute Difference between Unique Twitter Population and Census Block Population in Downtown San Diego in 2015 for both weekdays and weekends.

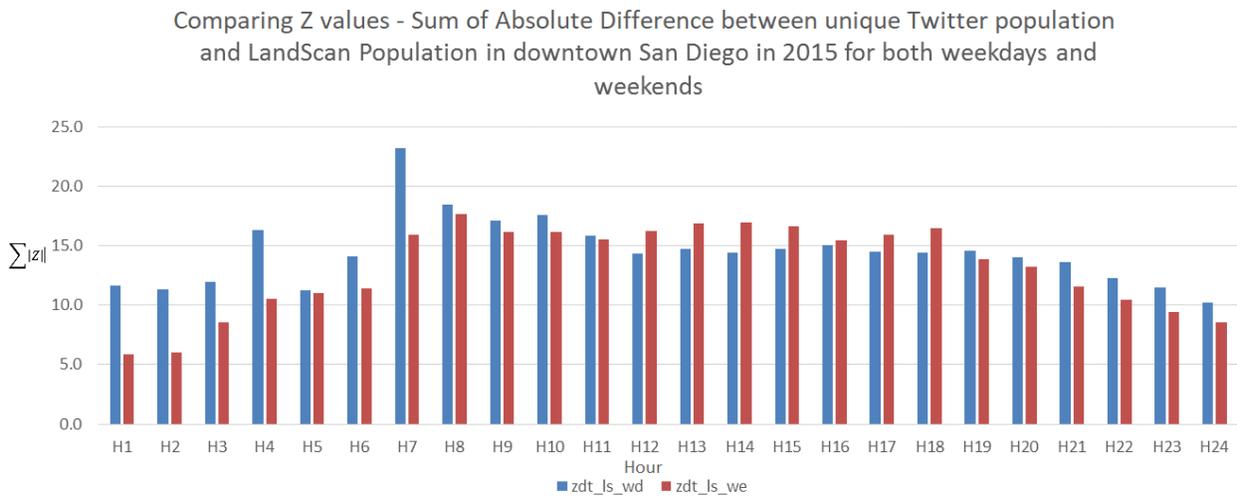

**Figure 11**. Comparing Z values – Sum of Absolute Difference between Unique Twitter Population and LandScan Population in Downtown San Diego in 2015 for both weekdays and weekends.

## 4. Transforming unique twitter users to estimated population with spatial and temporal variation factors

The previous sections illustrate how to calculate the dynamic changes of unique Twitter users in high spatial and temporal resolution units. The next step is to create a dynamic population model to transform the numbers of unique Twitter users to estimated population in order to match the need of building TAZ for evacuation planning. We proposed a simplified population estimation model using census blocks, land use data, and dasymetric mapping methods like the following:

$$\widehat{D}_{hx \cap a} = UserNumber_{hx \cap A} * (T_{hx}) * (S_{hx \cap A}) \qquad (2)$$

where:
  $hx$ = the time slot x;
  A = census block ID;
  $UserNumber_{hx \cap A}$ = the count of original unique Twitter population in census block A during time slot $hx$;
  $T_{hx}$ = T-Value for certain time slot $hx$;
  $S_{hx \cap A}$ = Spatial Change Factor for Twitter population in census block A during time slot $hx$.

### 4.1 Temporal variation factor (t-value)

The temporal variation factor (T-value) is defined as a value of factor multiples with the frequency number of hourly average Twitter user in each census block or land use polygon. A temporal factor was based on hourly frequency changes of unique Twitter users within the County of San Diego. Table 3 illustrated the creation of temporal variation factor (T-value). First of all, we calculate the total number of unique Twitter users in the whole San Diego County at each hour (from 0am, 1am, 2am …). Then we select the highest number (at 18:00, 75690) as the base number (T-value = 1). Each T-value is calculated using the base number (75690) divided by the total unique Twitter user numbers in each time slot. For example, the T-value at 4am will be 75690 / 5481 = 13.81.

| Time Slot | Total Unique Twitter User | *T-value |
|---|---|---|
| 0:00 - 0:59 | 19836 | 3.82 |
| 1:00 - 1:59 | 10701 | 7.07 |
| 2:00 - 2:59 | 6003 | 12.61 |
| 3:00 - 3:59 | 4437 | 17.06 |
| 4:00 - 4:59 | 5481 | 13.81 |
| 5:00 - 5:59 | 12267 | 6.17 |
| 6:00 - 6:59 | 22707 | 3.33 |
| 7:00 - 7:59 | 31059 | 2.44 |
| 8:00 - 8:59 | 38367 | 1.97 |
| 9:00 - 9:59 | 44370 | 1.71 |
| 10:00 - 10:59 | 48285 | 1.57 |
| 11:00 - 11:59 | 52200 | 1.45 |
| 12:00 - 12:59 | 54027 | 1.40 |
| 13:00 - 13:59 | 54288 | 1.39 |
| 14:00 - 14:59 | 54549 | 1.39 |
| 15:00 - 15:59 | 57942 | 1.31 |
| 16:00 - 16:59 | 64206 | 1.18 |
| 17:00 - 17:59 | 71514 | 1.06 |
| 18:00 - 18:59 | **75690** | **1.00** |
| 19:00 - 19:59 | 74646 | 1.01 |
| 20:00 - 20:59 | 74124 | 1.02 |
| 21:00 - 21:59 | 67338 | 1.12 |
| 22:00 - 22:59 | 53244 | 1.42 |
| 23:00 - 23:59 | 36018 | 2.10 |

**Table 3.** The total unique Twitter user numbers in each time slot and their T-values.

Figure 12 shows the original unique Twitter user density map (a) and the estimated population density map (b) with temporal variation factor (T-value = 3.82) from 0:00 to 0:59 in San Diego downtown for Weekdays in 2015.

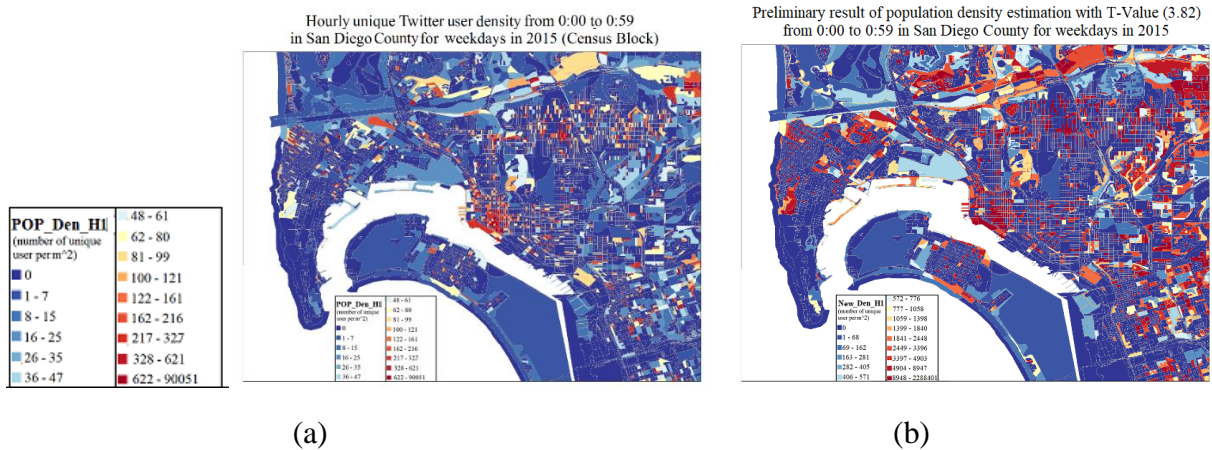

(a)            (b)

**Figure 12.** The original unique Twitter user density map (a) and the population density estimation (b) with temporal variation factor (T-value = 3.82) from 0:00 to 0:59 in San Diego County during weekdays in 2015.

**4.2** *Spatial change factor using dasymetric mapping methods (s-value).*

We utilized dasymetric mapping technique to redistribute the unique Twitter user population based on the ratio of average census population and the average hourly unique Twitter user population in each type of land use categories. The goal is to refine the population density maps based on different types of land use data (residential areas, commercial areas, etc.) and census data. The 2017 parcel land use data is downloaded from the San Diego Association of Governments (SANDAG) website (http://www.sandag.org).

    The census block boundaries (43,326 polygons in San Diego County) were overlaid with the 2016 parcel land use data (189,635 polygons) which created a union map with 670,913 polygons. The parcel land use data contains 10 types of land use which include unzoned, single family, minor multiple, restricted multiple, multiple residential, restricted commercial, commercial, industrial, agricultural, and special. We downgraded the 10 types of land cover into 6 categories which are unzoned, residential, commercial, industrial, agricultural, and special. The road section were added into the parcel shapefile by extracting the road polygons from SANDAG's land use shapefile which shares the same dimension with parcel data. The new land use map ended up with 7 types of land use in total (Table 2). Both census population and unique Twitter user population are re-distributed from the larger census block polygon to the finer polygons (subareas) in the overlaid map. The following formula (3) were applied to calculate the number of census population with certain land use type ($a$) as:

$$\widehat{SCP}_a = CP_A \left(\frac{SA_{A(a)}}{A_A}\right) \tag{3}$$

where:

$\widehat{SCP}_a$ = the estimated count of census population in subarea of land use $a$;
$CP_A$ = the count of census population in census block A;
$SA_{A(a)}$ = the area of subarea $a$ under census block A;
$A_A$ = the area of census block A;
$a$ = the land use type;
$A$ = census block ID.

The method of calculating unique Twitter population (formula 3) is similar to the way of re-distributing census population, while adding the temporal variation variable (T-value) into consideration. The count of unique Twitter population in census block A during time slot $hx$ ($TP_{hx \cap A}$) is acquired by multiplying average unique Twitter user with T-Value as:

$$TP_{hx \cap A} = tp_{hx \cap A}(T_{hx}) \tag{4}$$

where:

$tp_{hx \cap A}$ = the count of original Twitter population in census block A during time slot $hx$;
$T_{hx}$ = T-Value for certain time slot $hx$.

The estimated count of unique Twitter population in each subarea is then calculated based on the ratio of the size of subarea and area of census block A.

$$\widehat{STP}_{hx \cap a} = TP_{hx \cap A} \left(\frac{SA_{A(a)}}{A_A}\right) \tag{5}$$

where:

$\widehat{STP}_{hx \cap a}$ = the estimated count of unique Twitter population during time slot $hx$ in subarea of land use A;
$TP_{hx \cap A}$ = the count of unique Twitter population in census block A during time slot $hx$.

The estimated population density ($\widehat{D}_{hx \cap a}$) aims to estimate the hourly human population based on the ratio of the sum of census population in land use Type a and the sum of hourly unique Twitter user population in land use Type a. The ratio ($R_A$) is defined as:

$$R_A = \frac{\sum \widehat{SCP}_a}{\sum \widehat{STP}_{hx \cap a}} \tag{6}$$

$$\widehat{D}_{hx \cap a} = R_A \left(\frac{\widehat{STP}_{hx \cap a}}{SA_{A(a)}}\right) \tag{7}$$

while the estimated population density ($\widehat{D}_{hx \cap a}$) for certain land use type is the estimated count of unique Twitter population with $R$ and divided by the size of the corresponding subarea as formula (7).

Table 4 shows the area of 7 land use types in square kilometer, the total number of estimated unique Twitter population during H7 (6:00 to 6:59) and H21 (20:00 to 20:59) after applying T-value, and the estimated census population based on different types of land use.

| LC | Landuse | LC_area (km²) | twepop_h7 | twepop_h21 | cenpop |
|---|---|---|---|---|---|
| 0 | Unzoned | 6437.72 | 65.61 | 46.37 | 99553.74 |
| 1 | Residential | 1626.62 | 96.56 | 109.85 | 1128499.76 |
| 2 | Commerical | 394.29 | 42.07 | 49.32 | 112381.49 |
| 3 | Industrial | 322.65 | 21.50 | 18.15 | 24372.34 |
| 4 | Agricultrual | 1704.09 | 2.97 | 2.73 | 15602.81 |
| 5 | Special | 291.88 | 8.43 | 7.13 | 24342.04 |
| 6 | Road | 285.68 | 52.55 | 55.82 | 392448.08 |

**Table 4.** The area of 7 types of land use, the total number of estimated unique Twitter user population during 6:00 to 6:59 (twepop_h7) and 20:00 to 20:59 (twepop_h21), and the total number of estimated census population (cenpop) based on land use

Table 5 shows the ratio ($R_A$) which was calculated based on the division of cenpop ($\sum \widehat{SCP}_a$) with twepop_h7 ($\sum \widehat{STP}_{h7 \cap a}$) and twepop_h21 ($\sum \widehat{STP}_{h21 \cap a}$).

| LC | Landuse | ratio_h7 | ratio_h21 |
|---|---|---|---|
| 0 | Unzoned | 1517.25 | 2147.05 |
| 1 | Residential | 11686.73 | 10272.84 |
| 2 | Commerical | 2671.32 | 2278.48 |
| 3 | Industrial | 1133.52 | 1342.97 |
| 4 | Agricultrual | 5252.44 | 5708.23 |
| 5 | Special | 2889.01 | 3413.78 |
| 6 | Road | 7467.56 | 7030.45 |

**Table 5.** The Ratio for estimating the h7 (6:00 to 6:59) and h21 (20:00 to 20:59) real population and its corresponding land use type.

Figures 13 and 14 show the preliminary result of applying Equations (4) and (5) to adjust and re-distribute hourly unique Twitter user population into estimated population density. Based on the side by side comparison of estimated population density and the original unique Twitter user population, the estimated population is exaggerated and more population is redistributed on the residential area than the rest 6 types of land use due to the influence brought by census block data. Since census block data represents the count of population at home, it might improve the Twitter density maps to adjust the shortage of the people who may not tweet much when they are sleeping or at home. Figure 13 (a) shows more population in residential area instead of the original situation where downtown areas have higher density population.

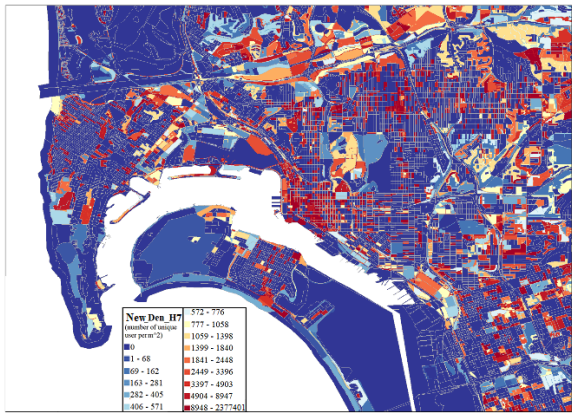
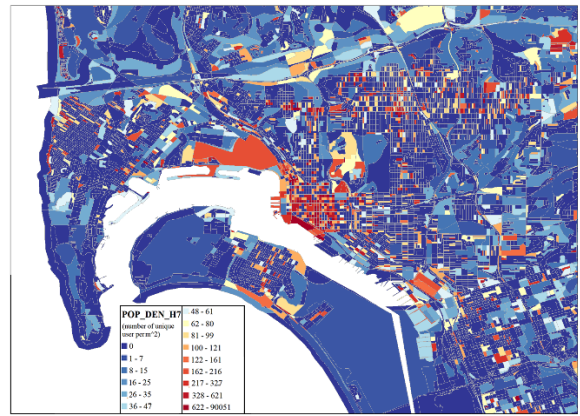

(a)  (b)

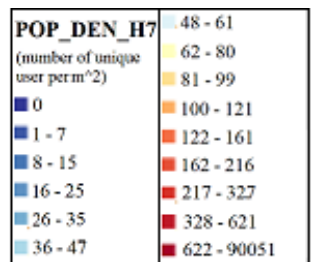

**Figure 13.** (a) Population density estimation with spatial variation factor and the dasymetric mapping method from 6:00 to 6:59 in San Diego downtown areas during Weekdays in 2015; (b) the original hourly unique Twitter user density from 6:00 to 6:59 in San Diego downtown areas during Weekdays in 2015.

In Figure 14, the maps show the comparison of the estimated population density map (a) and the original unique Twitter user population density map (b) from 20:00 to 20:59 during weekdays in San Diego downtown areas, with the 2010 population density based on 2010 census data (c). The result shows that dasymetric mapping technique might be able to provide a balanced population estimation comparing to the hourly unique Twitter user density and the census (night-time only) population. Comparing to the Twitter density map in the same time slot

(from 20:00 to 20:59), high population density areas, such as Balboa Park and San Diego Zoo, shopping malls, and San Diego International Airport, are adjusted based on their land use characteristics.

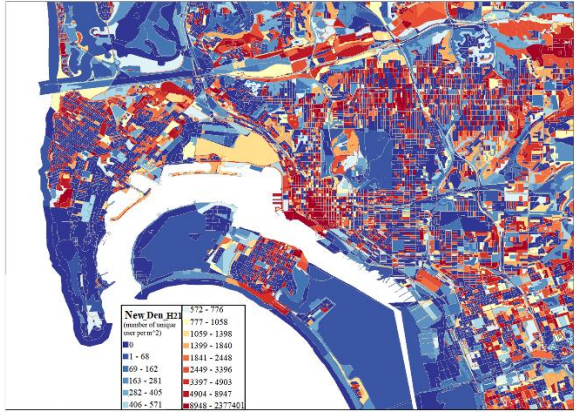
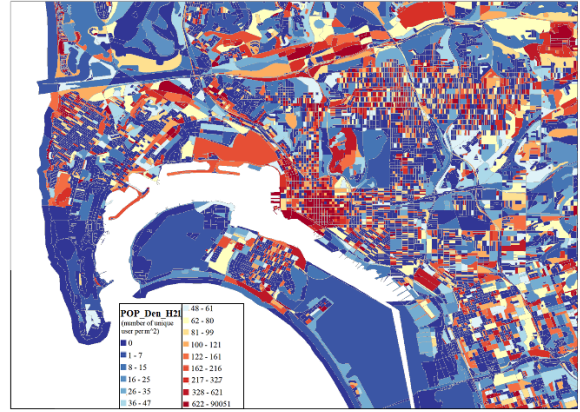

(a)            (b)

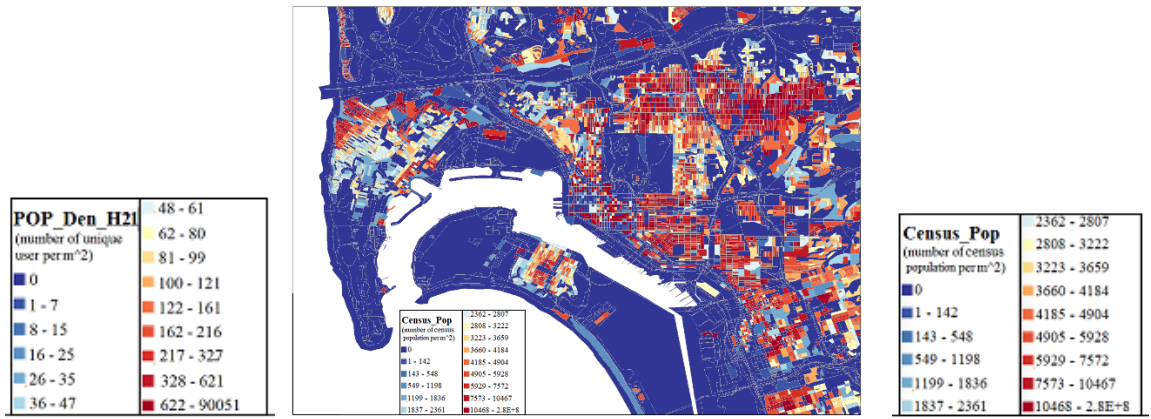

(c)

**Figure 14.** (a) Population density estimation map with the spatial variation factor and dasymetric mapping method from 20:00 to 20:59 in San Diego downtown areas during Weekdays in 2015.; (b) the original hourly unique Twitter user density map (middle) from 20:00 to 20:59 in San Diego downtown areas during Weekdays in 2015; (c) the 2010 census block population density map using census data.

## 5. Limitations and the future study

There are several research limitations in our study as the following:

a) Geo-tagged Twitter users can not represent the total population. In general, social media users are younger comparing to the general population, and more users live in urban areas than rural areas (Duggan et al. 2015). Although this study tried to introduce the temporal and spatial variation factor in our population model to compensate this limitation, we have not been able to validate these variables with real world data and actual population distribution.
b) It is very difficult to validate our dynamic population model because there is no similar data existed in San Diego County. We can only estimate the night time population to compare to the actual 2010 census data and the 2014 LandScan data. However, these data are not created original for displaying the dynamic hourly population density and may be suitable for the validation purpose.
c) Spatial and temporal factors in population estimation are usually correlated and should be considered together (Li et al. 2015). Our simplified model does not consider the autocorrelation between the spatial and temporal factors.
d) This study only utilizes one single social media data (Twitter) among many popular social media sites. In the future study, we should combine other social media, such as Instagram, Facebook check-in, Foursquare, and other possible digital footprints to enhance our population model. However, different types of social media platforms and digital footprints may have different types of spatiotemporal patterns, which will be another challenge research question.
e) The public Streaming APIs provided by Twitter is not very stable. We found that unequal number of tweets collect in different months and days, which may create some biases in our estimation of population density. For example, the Twitter use activities during March and April may more influence to the final population estimation result.

To improve and refine our future study of population density models, we are planning to use more complicated dasymetric mapping methods similar to intelligent dasymetric mapping technique (IDM) (Mennis and Hultgren 2006) to calculate the probability of population distribution in a more detailed land use category and census blocks using other spatial statistic methods, such as Weighted Linear Combination (WLC).

We recognized that validation is a key challenge to evaluate our dynamic population estimation model. While collecting dynamic population from real world in a large area is extremely difficult, it might be possible to partially compare the estimate during a certain temporal duration with existing data. For example, Census American Community Survey (ACS) provides a daytime population estimate (McKenzie et.al. 2010). Therefore, we can measure the goodness of fit between the estimates from the model and ACS during daytime (e.g., 9 am to 3 pm, a core work hour). However, it is necessary to carefully consider the validation process since social media data are drawn from potentially biased population and the data may include not only local residents but also visitors whereas ACS data account for residents and workers. Taking visitors in San Diego into consideration is helpful for revealing the real pattern of human dynamic. Therefore, further social media data filtering procedures should be applied to identify local residents for validation. The finalized framework, with frontend web design and backend database, can be applied with real-time data as well in the future by upgrading the current 1 hour temporal resolution to 10 minutes or even higher scale.

To summarize, although the Twitter data cannot perfectly represent the entire population, this study has revealed the potential research framework using social media data to calculate dynamic change of population distribution patterns. The combination of multiple social media data, mobile phone records, and other digital footprints created by human beings will be a great source to study human dynamics and help us to understand different types of human behaviors, movements, and activities in high spatial and temporal resolution. This integration of utilizing multiple sources of information would be able to increase the demographic comprehensiveness of this research. These information can facilitate the improvement of our transportation systems, emergency evacuation procedures, and urban planning in the future.

**Acknowledgements**

This material is based upon work supported by the National Science Foundation under Grant No. 1634641, IMEE project titled "Integrated Stage-Based Evacuation with Social Perception Analysis and Dynamic Population Estimation" and Grant No. 1416509, IBSS project titled "Spatiotemporal Modeling of Human Dynamics Across Social Media and Social Networks". Any opinions, findings, and conclusions or recommendations expressed in this material are those of the author and do not necessarily reflect the views of the National Science Foundation.